# Syndrome Decoding of Reed–Solomon Codes Beyond Half the Minimum Distance based on Shift-Register Synthesis

Georg Schmidt, *Student Member, IEEE*, Vladimir R. Sidorenko, *Member, IEEE*, and
Martin Bossert *Senior Member, IEEE*
Department of Telecommunications and Applied Information Theory, University of Ulm, Germany
{georg.schmidt,vladimir.sidorenko,martin.bossert}@uni-ulm.de

*Abstract*— In this paper, a new approach for decoding low–rate Reed–Solomon codes beyond half the minimum distance is considered and analyzed. Unlike the Sudan algorithm published in 1997, this new approach is based on multi–sequence shift–register synthesis, which makes it easy to understand and simple to implement. The computational complexity of this shift–register based algorithm is of the same order as the complexity of the well–known Berlekamp–Massey algorithm. Moreover, the error correcting radius coincides with the error correcting radius of the original Sudan algorithm, and the practical decoding performance observed on a $q$-ary symmetric channel (QSC) is virtually identical to the decoding performance of the Sudan algorithm. Bounds for the failure and error probability as well as for the QSC decoding performance of the new algorithm are derived, and the performance is illustrated by means of examples.

*Index Terms*— Reed–Solomon codes, decoding beyond half the minimum distance, Interleaved Reed–Solomon codes, multi–sequence shift–register synthesis,

## I. INTRODUCTION

Reed–Solomon codes are usually decoded algebraically by *Bounded Minimum Distance* (BMD) decoders which are capable of decoding all error patterns up to half the minimum distance, and no error patterns beyond this limit. Such classical BMD decoding approaches are mostly syndrome based, i.e., they work by creating a *syndrome* from the received word, which allows to uniquely reconstruct the error pattern, provided that its weight is not larger than half the minimum distance of the code. There have been several straight forward approaches like [1] and [2] for extending these syndrome based techniques for decoding errors beyond half the minimum distance using an extensive post processing step. However, for complexity reasons the decoding radius can usually only be slightly increased using such techniques.

When the original *Sudan algorithm* was published in 1997 [3], it attracted a lot of interest among experts, since it allows a Reed–Solomon code of rate $R \lesssim \frac{1}{3}$ to be decoded beyond half the minimum distance using algebraic methods.

The research work of Vladimir R. Sidorenko and Georg Schmidt is supported by Deutsche Forschungsgemeinschaft (DFG), Germany, under project BO 867/14 and project BO 867/15 respectively. Some of the results presented in this paper have also been presented on the 2006 IEEE International Symposium on Information Theory in Seattle, USA. This work has been submitted to the IEEE for possible publication. Copyright may be transferred without notice, after which this version will be superseded.

More precisely, the Sudan algorithm considers the problem of decoding a Reed–Solomon code as bivariate polynomial interpolation and polynomial factorization problem, which can be solved in polynomial time.

In [4] we presented a syndrome based decoding method, which uses multi–sequence shift–register synthesis to decode Reed–Solomon codes of rate $R \lesssim \frac{1}{3}$ beyond half the minimum distance. In this paper, we analyze this decoding method and demonstrate, that it practically gives the same decoding performance as the original Sudan algorithm [3].

## II. BMD DECODING OF REED–SOLOMON CODES

Unlike the Sudan algorithm which is based on polynomial interpolation, our new decoding approach aims at utilizing classical syndrome based methods. More precisely, we use an approach based on multi–sequence shift–register synthesis for locating the erroneous positions in the received words. To ease understanding of this approach, and to introduce the notation, we start with a brief overview over classical syndrome based BMD decoding of Reed–Solomon codes based on shift–register synthesis.

*Reed–Solomon Codes*

For our purposes, it is most convenient to define Reed–Solomon codes in the frequency domain, i.e., by the spectra of their codewords. Roughly speaking, a spectrum of a Reed–Solomon codeword is "band–limited", i.e., it possesses at most $k$ non-zero coefficients and at least $n-k$ cyclically consecutive zero coefficients. Terms like "spectrum" or "band–limited" are usually related to the Fourier Transform. To use such terms in the context of coding theory, we start with defining the *Discrete Fourier Transform* (DFT) over the field $\mathbb{F}_q$:

**Definition 1 (Discrete Fourier Transform in $\mathbb{F}_q$)**
Let $p(x) = p_0 + p_1 x + \cdots + p_{n-1} x^{n-1}$ be a polynomial of degree $\deg(p(x)) \leq n-1$ over $\mathbb{F}_q$. Further, let $\alpha \in \mathbb{F}_q$ be some element of order $n$. Then, the polynomial

$$P(x) = \mathscr{F}(p(x)) = P_0 + P_1 x + \cdots + P_{n-1} x^{n-1}$$

whose coefficients are calculated by $P_i = p(\alpha^i)$ is called the Fourier Transform of $p(x)$ over the field $\mathbb{F}_q$.

The inverse discrete Fourier Transform is calculated by

$$p(x) = \mathscr{F}^{-1}(P(x)) = p_0 + p_1 x + \cdots + p_{n-1} x^{n-1},$$

where $P(x)$ is a polynomial of $\deg(P(x)) \leq n-1$, and $p_i = n^{-1} P(\alpha^{-i})$. In the terminology of the Fourier Transform we call $p(x)$ *time domain polynomial* and $P(x)$ the *frequency domain polynomial* or the *spectrum* of $p(x)$.

Definition 1 enables us to describe Reed–Solomon codes by means of their codeword spectra. As already mentioned, the spectrum of a codeword of a Reed–Solomon code is characterized by $n-k$ cyclically consecutive zero coefficients. For our purposes, it is not necessary to define Reed–Solomon codes in their most general form. For the sake of simplicity, we restrict our considerations to codes, whose codeword spectra are zero in the $n-k$ rightmost positions. More precisely, we define a Reed–Solomon codes as follows:

**Definition 2 (Reed–Solomon code)**
*Let*

$$\{C(x)\} = \left\{ \sum_{i=0}^{k-1} C_i x^i,\ C_i \in \mathbb{F}_q \right\}$$

*be the set of all polynomials of $\deg(C(x)) < k$ with coefficients $C_i$ from $\mathbb{F}_q$, and let $\alpha \in \mathbb{F}_q$ be some element of order $n$. Then, a Reed–Solomon code $\mathcal{C} = \mathcal{RS}(q; n, k)$ of length $n$ and dimension $k$ can be defined as the set of polynomials*

$$\mathcal{C} = \left\{ c(x) = \mathscr{F}^{-1}(C(x)) \mid C(x) \in \{C(x)\} \right\}.$$

*The polynomials $c(x) = c_0 + c_1 x + \cdots + c_{n-1} x^{n-1}$, or equivalently the vectors $\boldsymbol{c} = (c_0, c_1, \ldots, c_{n-1})$, are called the codewords of the Reed–Solomon code.*

Due to the *Polynomial Factor Theorem*, a frequency polynomial $C(x)$ has at most $k-1$ roots. Hence, due to the properties of the DFT, a codeword has at most $k-1$ zero coefficients, which means, that each codeword has at least weight $n-k+1$. Thus, by the *Singleton bound*, the minimum distance of a Reed–Solomon code pursuant to Definition 2 is exactly $d = n-k+1$.

*BMD Decoding in the Frequency Domain*

Now, we assume that the codewords of $\mathcal{C}$ are transmitted over a noisy channel, which adds a random error polynomial $e(x) = e_0 + e_1 x + \cdots + e_{n-1} x^{n-1}$ with $t$ non-zero coefficients form $\mathbb{F}_q$. Hence, at the channel output we observe the word

$$y(x) = c(x) + e(x).$$

To decode the errors in this word, we calculate a *syndrome* from $y(x)$, which only depends on the error word $e(x)$ and not on the codeword $c(x)$. To do so, we utilize the spectral properties of a Reed–Solomon code. More precisely, we calculate the spectrum $Y(x) = \mathscr{F}(y(x)) = C(x) + E(x)$, which is the sum of the codeword spectrum $C(x) = \mathscr{F}(c(x))$ and the error spectrum $E(x) = \mathscr{F}(e(x))$. Then, we define $S_j = Y_{k+j} = E_{k+j}, j = 0, \ldots, n-k-1$. Since $\deg(C(x)) < k$, the sequence $\mathcal{S} = (S_0, S_1, \ldots, S_{n-k-1})$ only depends on the error specturm $E(x)$, and hence $\mathcal{S}$ acts as syndrome for decoding $y(x)$.

To locate the erroneous positions in $y(x)$, we use a technique introduced by Peterson [5] and define a polynomial $\lambda(x) = \lambda_0 + \lambda_1 x + \cdots + \lambda_{n-1} x^{n-1}$ such that $\lambda_j = 0$ whenever the corresponding coefficient $e_j$ of $e(x)$ is non-zero, and $\lambda_j \neq 0$, whenever $e_j = 0$. In this way, we have $\lambda_j \cdot e_j = 0\ \forall\ j = 0, \ldots, n-1$. Due to the well–known *convolution theorem* of the DFT, this relation is transformed into

$$\Lambda(x) \cdot E(x) \equiv 0 \mod x^n - 1. \quad (1)$$

The $t$ roots $\alpha^{-j_1}, \ldots, \alpha^{-j_t}$ of the frequency domain polynomial $\Lambda(x)$ correspond to the locations $j_1, \ldots, j_t$ of the erroneous positions in $y(x)$. Therefore

$$\Lambda(x) = \Lambda_0 + \Lambda_1 x + \cdots + \Lambda_t x^t$$

is called *error locator polynomial*. Since we are only interested in the roots of $\Lambda(x)$ and $\Lambda_0 \neq 0$, we are always able to normalize $\Lambda(x)$ w.l.o.g. such that $\Lambda_0 = 1$.

Equation (1) gives rise to a linear system of $n$ equations. From these equations, $n-k-t$ equations only depend on the $n-k$ coefficients from $E(x)$, which coincide with the elements $S_0, \ldots, S_{n-k-1}$ of the syndrome, and the unknown coefficients of the error locator polynomial $\Lambda(x)$. Hence, we extract a linear system of $n-k-t$ equations and $t$ unknowns. With $m = n-k$, this system of equations can be represented by the matrix equation

$$\underbrace{\begin{pmatrix} S_0 & S_1 & \ldots & S_{t-1} \\ S_1 & S_2 & \ldots & S_t \\ \vdots & \vdots & & \vdots \\ S_{m-t-1} & S_{m-t} & \ldots & S_{m-2} \end{pmatrix}}_{\boldsymbol{S}} \cdot \underbrace{\begin{pmatrix} \Lambda_t \\ \Lambda_{t-1} \\ \vdots \\ \Lambda_1 \end{pmatrix}}_{\boldsymbol{\Lambda}} = \underbrace{\begin{pmatrix} -S_t \\ -S_{t+1} \\ \vdots \\ -S_{m-1} \end{pmatrix}}_{\boldsymbol{T}}. \quad (2)$$

If (2) has a unique solution, it can be used to calculate the coefficients of a unique error locator polynomial $\Lambda(x)$. In this case, the locations of the erroneous symbols are uniquely determined. However, a unique solution can only exist, if the number of columns in $\boldsymbol{S}$ (number of unknowns) is not larger than the number of rows (number of equations), i.e., as long as

$$t \leq \left\lfloor \frac{n-k}{2} \right\rfloor.$$

Moreover, from the Reed–Solomon code properties we know that $\boldsymbol{S}$ has full rank and therefore a unique solution must always exists as long as the number of errors is not larger than half the minimum distance. Consequently, the BMD error correcting radius is specified by

$$\tau = \left\lfloor \frac{n-k}{2} \right\rfloor = \left\lfloor \frac{d-1}{2} \right\rfloor.$$

Carefully considering Equation 2 yields that the matrix $\boldsymbol{S}$ is a *Toeplitz* matrix. This allows us to rewrite (2) in the form of the linear recursion

$$S_j = -\sum_{h=1}^{t} \Lambda_h S_{j-h},\ j = t+1, \ldots, 2t. \quad (3)$$

Since this linear recursion can be realized by a linear feedback shift-register where the (unknown) coefficients of $\Lambda(x)$ are the feedback connection weights, the error locator polynomial $\Lambda(x)$ can be determined by synthesizing the shortest possible linear feedback shift-register capable of generating $\mathcal{S}$. This task is performed very efficiently by the well–known Berlekamp–Massey algorithm [6], [7]. If $y(x)$ is erroneous in $t$ positions, then the syndrome sequence has linear complexity $t$ due to *Blahut's Theorem* [1], [8]. Hence, the Berlekamp–Massey algorithm will synthesize a shift-register of length $t$. As long as $t \leq \lfloor \frac{n-k}{2} \rfloor$, (2) has only one solution, and consequently the Berlekamp–Massey algorithm always synthesizes a valid $\Lambda$-polynomial.

After a valid error locator polynomial $\Lambda(x)$ is determined, its roots $\alpha^{-j_1}, \ldots, \alpha^{-j_t}$ give rise to the positions $j_1, \ldots, j_t$ of the erroneous symbols in $y(x)$. To evaluate the errors at these positions, i.e., to determine the values of the symbols $e_{j_1}, \ldots, e_{j_t}$ we may use standard methods like the *Forney Algorithm* [9] or *Recursive Extension* [10].

## III. DECODING BEYOND HALF THE MINIMUM DISTANCE

Clearly, the BMD decoding strategy described in the previous section uses a spherical decoding region. To decode a code $\mathcal{C} \subset \mathbb{F}_q^n$ of length $n$ and minimum Hamming distance $d$, a BMD decoder surrounds each codeword $\boldsymbol{c} \in \mathcal{C}$ by a sphere of radius $t_{\max} \leq \tau = \lfloor \frac{d-1}{2} \rfloor$ as sketched in Fig. 1a. All vectors $\boldsymbol{y}$ within this sphere are decoded into $\boldsymbol{c}$. Since $t_{\max} \leq \tau$, the decoding spheres are not overlapping, and consequently each vector $\boldsymbol{y} \in \mathbb{F}_q^n$ is included either in exactly one sphere, or in no sphere at all. Hence, all received vectors $\boldsymbol{y}$ within a sphere around $\boldsymbol{c}$ are uniquely decoded into $\boldsymbol{c}$, and all vectors outside the spheres are not decodable. Thus, a BMD decoder either yields a unique decoding result (either correct or erroneous), or a *decoding failure*, i.e., no result at all.

In contrast to this, a *Maximum Likelihood* (ML) decoder is capable of decoding errors beyond half the minimum distance. For this purpose, it covers the complete space $\mathbb{F}_q^n$ by irregular shaped *Voronoi cells* as depicted in Fig. 1b. Roughly speaking, a Voronoi cell around a codeword $\boldsymbol{c} \in \mathcal{C}$ contains all vectors $\boldsymbol{y} \in \mathbb{F}_q^n$ whose distance to $\boldsymbol{c}$ is smaller than the distance to any other codeword in $\mathcal{C}$. Hence, an ML decoder covers the complete space $\mathbb{F}_q^n$, and consequently never yields a decoding failure. However, the complex shapes of the Voronoi cells are difficult to handle by algebraic methods, which means that known ML decoders generally possess a computational complexity, which grows exponentially in $n$.

To decode errors beyond half the minimum distance by algebraic methods, a common strategy is to stick to spherical decoding regions, but increase the maximum decoding radius $t_{\max}$ beyond half the minimum code distance. Such a strategy is commonly called *Bounded Distance* (BD) decoder. As illustrated in Fig. 1c, it is inevitable for BD decoders that some of the decoding spheres are overlapping. Consequently, the received words $\boldsymbol{y} \in \mathbb{F}_q^n$, which are included in more than one correcting sphere, cannot be uniquely decoded into a single codeword, since the codewords corresponding to all correcting spheres which contain the vector $\boldsymbol{y}$ are valid decoding results for the received word $\boldsymbol{y}$. In such situations, basically three strategies are conceivable:

1) The decoder selects one single codeword out of all valid decoding results
2) The decoder returns a list of all valid decoding results, that is, a list of all codewords which lie inside a sphere of radius $t_{\max}$ around the received word $\boldsymbol{y}$.
3) The decoder yields a decoding failure, i.e., it yields no solution at all

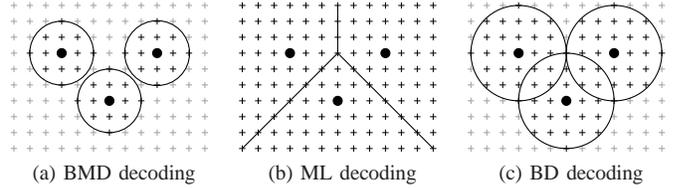

(a) BMD decoding     (b) ML decoding     (c) BD decoding

Fig. 1. Decoding regions for different decoding strategies.

The Sudan algorithm and its derivatives like [11], [12], etc., apply the second strategy and yield a list of codewords if the decoding result is not unique. If a unique solution is required, one candidate of this list has to be selected (usually the one with the smallest distance to the received word $\boldsymbol{y}$). In this case, the first strategy is applied. The third strategy is e.g. applied by [13], where the decoder may yield a decoding failure if multiple solutions exist.

If the volume of the overlapping decoding regions are small enough, the probability for a unique decoding result is quite close to one. In this case, all three decoding strategies practically result in a similar decoding performance, since there are only a very few cases in which the different strategies yield different decoding results.

We now describe a new decoding method with also applies the third strategy. We later demonstrate, that this approach practically yields the same QSC decoding performance as the classical Sudan algorithm [3].

### A. Virtual Extension to an Interleaved Reed–Solomon Code

*Interleaved Reed–Solomon* (IRS) codes like considered in [14], [15], [16], [17], [18], [19], [20], and others are capable of correcting error patterns beyond half the minimum distance. Usually, IRS code constructions are *homogeneous*, i.e., they consist of $l$ codewords of one single Reed–Solomon code. However, as suggested in [15] and investigated in [20], we also can construct *heterogeneous* IRS codes by taking the $l$ codewords from different Reed–Solomon codes with the same length but different dimensions. As shown in [20], decoding of both homogeneous and heterogeneous IRS codes can be performed very efficiently using multi–sequence shift–register synthesis.

Like proposed in [4], to decode a single Reed–Solomon code beyond half the minimum distance, we can virtually extend the code such that it forms a heterogeneous IRS code as described in [20]. To do this, we proceed as follows: Consider the polynomial

$$p(x) = p_0 + p_1 x + \cdots + p_{n-1} x^{n-1}$$

over $\mathbb{F}_q$ and denote by $p^{\langle i \rangle}(x)$ the polynomial

$$p^{\langle i \rangle}(x) = p_0^i + p_1^i x + \cdots + p_{n-1}^i x^{n-1}\,.$$

Consider the Reed–Solomon Code $\mathcal{C} = \mathcal{RS}(q; n, k)$, and define the codes

$$\mathcal{C}^{\langle i \rangle} = \left\{ c^{\langle i \rangle}(x) : c(x) \in \mathcal{C} \right\},\ i = 1, \ldots, l\,. \quad (4)$$

**Lemma 1** *Let $\mathcal{C} = \mathcal{RS}(q; n, k)$ be a Reed–Solomon code. If $i(k-1)+1 \leq n$, then each codeword from $\mathcal{C}^{\langle i \rangle}$ is a codeword of the Reed–Solomon code*

$$\mathcal{C}^{(i)} = \mathcal{RS}(q; n, i(k-1)+1)\,,$$

*of length $n$ and dimension $i(k-1)+1$, or in other words, $\mathcal{C}^{\langle i \rangle} \subseteq \mathcal{C}^{(i)}$.*

*Proof:* Every codeword $c^{\langle i \rangle}(x) \in \mathcal{C}^{\langle i \rangle}$ corresponds to a codeword $c(x) \in \mathcal{C}$, where $c(x) = \mathscr{F}^{-1}(C(x))$, and $\deg(C(x)) \leq k-1$. Due to the properties of the discrete Fourier Transform, the spectrum of $c^{\langle i \rangle}(x)$ is calculated by $\mathscr{F}\left(c^{\langle i \rangle}(x)\right) = C^i(x)$, where $\deg(C^i(x)) \leq i(k-1)$. Hence, by by Definition 2, the polynomial

$$c^{\langle i \rangle}(x) = \mathscr{F}^{-1}\left(C^i(x)\right)$$

is a codeword of a Reed–Solomon code of length $n$ and dimension $i(k-1)+1$. ■

To transform a codeword $c(x) = c_0 + c_1 x + \cdots + c_{n-1} x^{n-1}$ of a Reed–Solomon code $\mathcal{C} = \mathcal{RS}(q; n, k)$ into a codeword of a heterogeneous IRS code, we select a positive integer $l$, such that

$$l(k-1)+1 \leq n \quad (5)$$

is satisfied. Then, we create the words

$$\boldsymbol{c}^{\langle i \rangle} = \left(c_0^i, c_1^i, \ldots, c_{n-1}^i\right),\ i = 1, \ldots, l\,.$$

According to Lemma 1, we have

$$\boldsymbol{c}^{\langle i \rangle} \in \mathcal{C}^{(i)},\ i = 1, \ldots, l\,.$$

Hence, the $l \times n$ matrix

$$\boldsymbol{C} = \begin{pmatrix} \boldsymbol{c}^{\langle 1 \rangle} \\ \boldsymbol{c}^{\langle 2 \rangle} \\ \vdots \\ \boldsymbol{c}^{\langle l \rangle} \end{pmatrix} = \begin{pmatrix} c_0 & c_1 & \cdots & c_{n-1} \\ c_0^2 & c_1^2 & \cdots & c_{n-1}^2 \\ \vdots & \vdots & & \vdots \\ c_0^l & c_1^l & \cdots & c_{n-1}^l \end{pmatrix},$$

consisting of the row vectors $c^{\langle 1 \rangle}, c^{\langle 2 \rangle}, \ldots, c^{\langle l \rangle}$ can be considered as codeword of the heterogeneous IRS code $\mathcal{C}$ composed from the $l$ Reed–Solomon codes $\mathcal{C}^{(1)}, \mathcal{C}^{(2)}, \ldots, \mathcal{C}^{(l)}$.

Now, we assume that we transmit $c(x) \in \mathcal{C}$ over a channel, and observe the corrupted word $y(x) = c(x) + e(x)$ at the channel output. To reconstruct the transmitted codeword, we do not want to decode $y(x)$ with respect to $\mathcal{C}$, but rather process $y(x)$ in such a way that we are able to perform decoding with respect to the heterogeneous IRS code $\mathcal{C}$. For this purpose, we calculate the $l$ polynomials

$$y^{\langle i \rangle}(x) = y_0^i + y_1^i x + \cdots + y_{n-1}^i x^{n-1},\ i = 1, \ldots, l\,.$$

With the coefficients of these polynomials, we create the matrix

$$\boldsymbol{Y} = \begin{pmatrix} \boldsymbol{y}^{\langle 1 \rangle} \\ \boldsymbol{y}^{\langle 2 \rangle} \\ \vdots \\ \boldsymbol{y}^{\langle l \rangle} \end{pmatrix} = \begin{pmatrix} y_0 & y_1 & \cdots & y_{n-1} \\ y_0^2 & y_1^2 & \cdots & y_{n-1}^2 \\ \vdots & \vdots & & \vdots \\ y_0^l & y_1^l & \cdots & y_{n-1}^l \end{pmatrix}\,. \quad (6)$$

Roughly speaking, the matrix $\boldsymbol{Y}$ can be considered as "received" matrix of an IRS code. More precisely, $\boldsymbol{Y}$ possesses the following properties:

**Lemma 2** *Let $c(x) \in \mathcal{C}$ be a codeword of a Reed–Solomon code $\mathcal{C}$ according to Definition 2, transmitted over a noisy channel. Assume that the word $y(x) = c(x)+e(x)$ is received, and let $\boldsymbol{Y}$ be an $l \times n$ matrix composed from the coefficients of $y(x)$ pursuant to (6). Then, if $e(x) \equiv 0$, the matrix $\boldsymbol{Y}$ is a codeword of a heterogeneous IRS code composed of the $l$ Reed–Solomon codes*

$$\mathcal{C}^{(i)} = \mathcal{RS}(q; n, i(k-1)+1),\ i = 1, \ldots, l\,.$$

*Moreover, if $e(x)$ is a polynomial with $t$ non-zero coefficients $e_{j_1}, e_{j_2}, \ldots, e_{j_t}$, then the matrix $\boldsymbol{Y}$ is a corrupted IRS codeword with exactly $t$ erroneous columns at the positions $j_1, j_2, \ldots, j_t$.*

*Proof:* If $e(x) \equiv 0$, then $y(x) \in \mathcal{C}$, and by Lemma 1 and the explanations given above, we know that $\boldsymbol{Y}$ is a codeword of the heterogeneous IRS code $\mathcal{C}$. If $e(x)$ is a non-zero polynomial, the coefficients of the polynomial

$$y^{\langle i \rangle}(x) = y_0^i + y_1^i x + \cdots + y_{n-1}^i x^{n-1}$$

may be represented by

$$y_j^i = (c_j + e_j)^i \triangleq c_j^i + e_j^{[i]}\,.$$

Clearly, if $e_j = 0$, that is, if $j \notin \{j_1, j_2, \ldots, j_t\}$, then $e_j^{[i]} = 0\ \forall i = 1, \ldots, l$. On the other hand, if $e_j \neq 0$, that is, if $j \in \{j_1, j_2, \ldots, j_t\}$, then $e_j^{[i]}$ may be non-zero. Hence, the first symbol, and in most cases also the other symbols of the columns $j_1, j_2, \ldots, j_t$ in $\boldsymbol{Y}$ are erroneous, which means that there are exactly $t$ corrupted columns at the positions $j_1, j_2, \ldots, j_t$. ■

### B. Decoding by Syndrome Extension

Since due to Lemma 2 the matrix $\boldsymbol{Y}$ can be considered as corrupted codeword of a heterogeneous IRS code $\mathcal{C}$, the matrix $\boldsymbol{Y}$ may be decoded as IRS code word using the collaborative decoding strategies described in [20]. In other words, we are able to create the syndromes

$$\mathcal{S}^{[i]} = (S_0^{[i]}, \ldots, S_{n-i(k-1)-2}^{[i]}) = (Y_{i(k-1)+1}^{[i]}, \ldots, Y_{n-1}^{[i]}) \quad (7)$$

by taking the last $n - i(k-1) - 1$ coefficients from the spectra $Y^{[i]}(x) = \mathscr{F}\left(y^{[i]}(x)\right)$ for $i = 1, \ldots, l$. In this way, we obtain $l - 1$ auxiliary syndromes $\mathcal{S}^{[i]}$ with decreasing length in addition to the literal syndrome $\mathcal{S}^{[1]} = \mathcal{S}$ for the error word

$e(x)$. These syndromes are used to form the following system of equations:

$$\underbrace{\begin{pmatrix} \boldsymbol{S}^{[1]} \\ \boldsymbol{S}^{[2]} \\ \vdots \\ \boldsymbol{S}^{[l]} \end{pmatrix}}_{\boldsymbol{S}} \cdot \underbrace{\begin{pmatrix} \Lambda_t \\ \Lambda_{t-1} \\ \vdots \\ \Lambda_1 \end{pmatrix}}_{\boldsymbol{\Lambda}} = \underbrace{\begin{pmatrix} \boldsymbol{T}^{[1]} \\ \boldsymbol{T}^{[2]} \\ \vdots \\ \boldsymbol{T}^{[l]} \end{pmatrix}}_{\boldsymbol{T}} \quad (8)$$

where

$$\boldsymbol{S}^{[i]} = \begin{pmatrix} S_0^{[i]} & S_1^{[i]} & \ldots & S_{t-1}^{[i]} \\ S_1^{[i]} & S_2^{[i]} & \ldots & S_t^{[i]} \\ \vdots & \vdots & & \vdots \\ S_{m_i-t-1}^{[i]} & S_{m_i-t}^{[i]} & \ldots & S_{m_i-2}^{[i]} \end{pmatrix},$$

$$\boldsymbol{T}^{[i]} = \begin{pmatrix} -S_t^{[i]} \\ -S_{t+1}^{[i]} \\ \vdots \\ -S_{m_i-1}^{[i]} \end{pmatrix},$$

and $m_i = n - i(k-1) - 1$. In principle, $\Lambda(x)$ can now be found by solving (8) provided that $\boldsymbol{S}$ has full rank. This approach can be considered as generalization of the Peterson algorithm. Standard methods from linear algebra usually require a computational complexity of order $\mathcal{O}(n^3)$ for solving (8).

## IV. ERROR CORRECTING RADIUS AND THRESHOLD RATES

For ordinary IRS codes, the parameter $l$ is a freely selectable parameter, which determines the number of rows in a codeword matrix $\boldsymbol{C} \in \mathcal{C}$. For an IRS code obtained by syndrome extension, the parameter $l$ first of all has to fulfill (5). This means, that the maximum $l$ for syndrome extension is somehow limited by the rate $R = \frac{k}{n}$ of the original Reed–Solomon code $\mathcal{C}$. Moreover, to be able to use the $l$th auxiliary syndrome $\boldsymbol{S}^{[l]}$ in the system of equations pursuant to (8) with $t$ unknowns, $m_l > t$ has to be fulfilled. This restricts the maximum number of correctable errors for a given $l$. Hence, the maximum number of correctable errors is determined by the choice of $l$, which in turn is limited by the rate $R$ of the Reed–Solomon code $\mathcal{C}$.

In the following sections, we investigate the relation between the code rate $R$, the maximum error correcting radius $t_{\max}$, and the parameter $l$.

### A. Maximum Error Correcting Radius

Using only the literal syndrome $\mathcal{S} = \mathcal{S}^{[1]}$ to decode $\mathcal{C} = \mathcal{RS}(q; n, k)$, we are able to decode errors up to half the minimum distance, i.e., we obtain the error correcting radius

$$t_{\max}^{[1]} = t_{\max} = \left\lfloor \frac{n-k}{2} \right\rfloor .$$

By using $= \mathcal{S}^{[1]}$ together with an auxiliary syndrome $= \mathcal{S}^{[2]}$ of length $n - 2k + 1$, the system of equations (8) may have a unique solution as long as

$$t \leq 2n - 3k - 2t + 1 ,$$

which means that the maximum error correcting radius is increased to

$$t_{\max}^{[2]} = \left\lfloor \frac{2n - 3k + 1}{3} \right\rfloor . \quad (9)$$

Generally, by using the $l$ syndromes $\mathcal{S}^{[1]}, \mathcal{S}^{[2]}, \ldots, \mathcal{S}^{[l]}$, (8) may have a unique solution, as long as $t$ fulfills

$$t \leq ln - \frac{l(l+1)}{2}k + -lt + \frac{l(l-1)}{2} .$$

Consequently, the maximum error correcting radius is determined by

$$t_{\max}^{[l]} = \left\lfloor \frac{2ln - l(l+1)k + l(l-1)}{2(l+1)} \right\rfloor . \quad (10)$$

Equation (10) coincides with the error correcting radius of a heterogeneous IRS code composed of the $l$ Reed–Solomon codes $\mathcal{C}^{(1)}, \mathcal{C}^{(2)}, \ldots, \mathcal{C}^{(l)}$ as described in [20].

### B. Threshold Rates

To understand, how the choice of $l$ is restricted by the rate of $\mathcal{C}$, we consider the spectrum $Y(x) = \mathscr{F}(y(x))$ of the received word $y(x) = c(x) + e(x)$ and the spectrum of the auxiliary polynomial $y^{\langle 2 \rangle}(x)$. As explained in Section II, each codeword $c(x) \in \mathcal{C}$ has $n - k$ consecutive zero coefficients in its spectrum, and therefore there exist $n - k$ consecutive coefficients in $Y(x)$ which only depend on $e(x)$. Informally speaking, the spectrum of $y(x)$ possesses a "window" of size $n - k$ through which we are able to observe $n - k$ coefficients of the transformed error $E(x) = \mathscr{F}(e(x))$. As depicted in Fig. 2a, these coefficients constitute the literal syndrome $\mathcal{S} = \mathcal{S}^{[1]}$. Now, we consider the spectrum

$$Y^{\langle 2 \rangle}(x) = Y^2(x) = (C(x) + E(x))^2 = C^2(x) + E^{[2]}(x) .$$

The spectral bandwidth of $Y^2(x)$ is $k - 1$ elements larger than the bandwidth of the original codeword spectrum $Y(x)$. Thus, as illustrated in Fig. 2b, the size of the "window" in the spectrum $Y^{\langle 2 \rangle}(x)$ is reduced by $k - 1$ coefficients, which results in an auxiliary syndrome $\mathcal{S}^{[2]}$ of length $n - 2k + 1$.

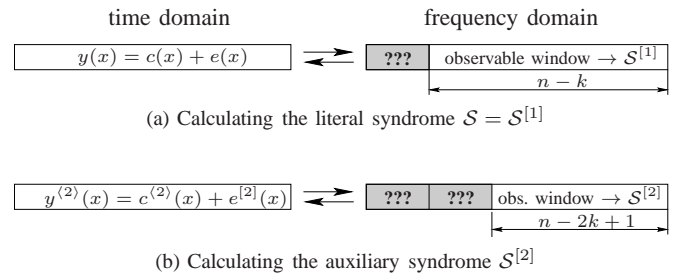

(a) Calculating the literal syndrome $\mathcal{S} = \mathcal{S}^{[1]}$

(b) Calculating the auxiliary syndrome $\mathcal{S}^{[2]}$

Fig. 2. Rate restrictions with virtual Syndrome Extension.

Clearly, if $R \gtrsim 1/2$, the "window" in the auxiliary spectrum $Y^{[2]}(x) = \mathscr{F}(y^{[2]}(x))$ is completely closed, which means that we cannot obtain a non-empty auxiliary syndrome $\mathcal{S}^{[2]}$. Moreover, even if there exists a non-empty syndrome $\mathcal{S}^{[2]}$, this syndrome can only be used to extend the error correcting radius if its length is greater than $\lfloor \frac{n-k}{2} \rfloor + 1$. In other words, syndrome extension can only be used if

$$n - 2k + 1 > \left\lfloor \frac{n-k}{2} \right\rfloor + 1$$

is satisfied. Solving this inequality for $R = k/n$ yields that the error correcting radius can only be extended for codes of rate $R \lesssim 1/3$. Hence, exactly like the classical Sudan algorithm [3], our syndrome extension procedure is only suitable for low–rate Reed–Solomon codes.

Generally, the decreasing lengths of the auxiliary syndromes $\mathcal{S}^{[2]}, \ldots, \mathcal{S}^{[l]}$ induce a restriction on the maximum rate of a code, for which $l$ syndromes effectively can be used. More precisely, to apply $l > 1$ syndromes, the code rate has to be small enough such that

$$t_{\max}^{[l-1]} + 2 \leq n - l(k-1) - 1 \quad (11)$$

is satisfied. For a fixed parameter $l$ we call the maximum rate of code $\mathcal{C}$ of length $n$ for which (11) is satisfied the *lth threshold rate* of $\mathcal{C}$. The other way round, for a given Reed–Solomon code of length $n$ and rate $R = k/n$, the parameter $l$ should be chosen to be the largest positive integer for which (11) is satisfied in order to maximize the error correcting radius. If (11) cannot be satisfied for $l > 1$, then no auxiliary syndrome can be created and the maximum correcting radius is limited by half the minimum distance.

A closed form approximation for the $l$th threshold rate can be obtained by neglecting the floor operation in (10). Inserting (10) into (11) and skipping $\lfloor \cdot \rfloor$ yields

$$R_{\text{th}}^{(l)} = \frac{2}{l(l+1)} + \frac{l^2 - 3l - 2}{nl(l+1)} \approx \frac{2}{l(l+1)}, \quad (12)$$

where $R_{\text{th}}^{(l)}$ is the rate below which we are always able to use the $l$ auxiliary syndromes $\mathcal{S}^{[2]}, \ldots, \mathcal{S}^{[l]}$, and consequently obtain the error correcting radius $t_{\max}^{[l]}$. Vice versa, for a given code $\mathcal{C}$ of length $n$, dimension $k > 1$, and $R \leq R_{\text{th}}^{(l)}$, we are able to solve (12) for $l$, and find that the number of generable syndromes is at least

$$l = \left\lfloor \frac{\sqrt{(Rn+3)^2 + 8(Rn-1)(n-1)} - (Rn+3)}{2(Rn-1)} \right\rfloor. \quad (13)$$

Note, that since we neglect the $\lfloor \cdot \rfloor$ operation for calculating $R_{\text{th}}^{(l)}$, there might be cases (particularly for small code lengths) for which we are able use $l$ syndromes even for code rates slightly higher than $R_{\text{th}}^{(l)}$.

The error correcting radius calculated by (10) and the threshold rates specified by (12) largely coincide with the error correcting radii and the threshold rates realized by the classical Sudan Algorithm [3]. Moreover, there are further relations between our syndrome extension algorithm and the Sudan algorithm. In [12], Roth and Ruckenstein proposed an algorithm for efficiently calculating the $Q$–polynomial required by the Sudan algorithm. This algorithm applies an iterative procedure, which synthesizes a bivariate polynomial $Q^*(x, y)$ from a bivariate syndrome $S(x, y) = \sum_{h=1}^{l} S^{(h)}(x) y^i$. The coefficients $S^{(h)}(x)$ of this polynomial are equivalent to the syndrome sequences $\mathcal{S}^{[h]}$, which we use to state (8). However, Roth and Ruckenstein use their algorithm for synthesizing a $Q$–polynomial, from which the valid decoding solutions have to be extracted by a recursive reconstruction algorithm. In contrast to this, we simply use the syndromes $\mathcal{S}^{[1]}, \ldots, \mathcal{S}^{[l]}$ for determining the error locator polynomial $\Lambda(x)$.

There exists another Sudan based algorithm proposed in [13]. This algorithm also does not create a list in ambiguous situations, but fails if no unique solution can be obtained. Unlike our algorithm, the algorithm from [13] is based on a polynomial interpolation approach, which aims in directly reconstructing $C(x)$, rather than using a two step strategy, first calculating an error locator polynomial and then evaluating the erroneous symbols. Moreover, the algorithm in [13] requires a matrix inversion, which has a computational complexity in the order of $\mathcal{O}(n^3)$, when performed by Gaussian elimination.

We now describe a method for synthesizing $\Lambda(x)$, which is based on multi–sequence shift–register synthesis. Our algorithm requires simply $l$ times the complexity of the Berlekamp–Massey algorithm, which is known to be not larger than $\mathcal{O}(n^2)$ (see e.g. [21]). Hence, the computational complexity of our algorithm is in the order of $\mathcal{O}(n^2)$.

## V. Efficient Error Location by Multiple Shift-Register Synthesis

For classical BMD decoding as described in Section II, the linear system of equations (2) can be reformulated into a linear shift–register synthesis problem which can efficiently be solved by the Berlekamp–Massey algorithm. As described in [20], the same principle can also be applied for decoding IRS codes. This means, we can reformulate (8) into the following set of $l$ linear recursions:

$$S_j^{(i)} = -\sum_{h=1}^{t} \Lambda_h S_{j-h}^{(i)}, j = t, \ldots, n-k-1, \ i = 1, \ldots, l. \quad (14)$$

For homogeneous IRS codes, the $l$ syndrome sequences $S_j^{(1)}, S_j^{(2)}, \ldots, S_j^{(l)}$ involved in this set of linear recursions all have the same length. In this case (14) could be solved by the *Feng–Tzeng algorithm* [22], [23]. However, the syndrome extension method explained above generally yields heterogeneous IRS codes. In this case, the lengths of the syndrome sequences $S_j^{(1)}, S_j^{(2)}, \ldots, S_j^{(l)}$ are different. As shown in [24], [25], the Feng–Tzeng algorithm does not always yield a correct solution in this case. Hence, we use the modified multi–sequence shift–register synthesis algorithm proposed in [24], [25]. A description of this algorithm adapted for our purposes is given by Algorithm 1. If we apply this algorithm to $l$ syndrome sequences $\mathcal{S}^{[1]}, \ldots, \mathcal{S}^{[l]}$, its computational complexity will be smaller than $l$ times the complexity of the Berlekamp–Massey algorithm.

If (8) does not have a unique solution, the polynomial $\Lambda(x)$ may not be a valid error locator polynomial. To be able to detect this, we consider the following definition:

**Definition 3 ($t$-valid $\Lambda$–polynomial)**
*A polynomial $\Lambda(x)$ over $\mathbb{F}_q$ is called $t$-valid if it is a polynomial of degree $t$ and possesses exactly $t$ distinct roots in $\mathbb{F}_q$.*

```
input: 𝒮^[h] = {S_i^[h]}_{i=0}^{n-h(k-1)-2}, h = 1, ..., l
begin
    t ← 0, Λ(x) ← 1, m ← n - k
    m_h ← length(𝒮^[h]), M^(h) ← m - m_h,
    t^(h) ← 0, d^(h) ← 1, Λ^(h)(x) ← 0, for h = 1, ..., l
    for each M from 1 to m do
        for each h from 1 to l do
            if M - t > m - m_h then
                d ← S^(h)_{M-m+m_h} + Σ_{j=1}^{t} Λ_j S^(h)_{M-j-m+m_h}
                if d ≠ 0 then
                    if M - t ≤ M^(h) - t^(h) then
                        Λ(x) ← Λ(x) - (d/d^(h)) Λ^(h)(x) x^{M-M^(h)}
                    else
                        t̃ ← t, Λ̃(x) ← Λ(x)
                        Λ(x) ← Λ(x) - (d/d^(h)) Λ^(h)(x) x^{M-M^(h)}
                        t ← M - (M^(h) - t^(h))
                        t^(h) ← t̃, Λ^(h)(x) ← Λ̃(x)
                        d^(h) ← d, M^(h) ← M
end
output: t, Λ(x)
```

Algorithm 1.  Efficient Error Location Algorithm

By definition of the error locator polynomial, a polynomial $\Lambda(x)$ can only be a valid error locator polynomial, if it possesses exactly $t$ distinct roots $\mathbb{F}_q$. Hence, we accept a $\Lambda$–polynomial obtained from Algorithm 1 only if it complies with Definition 3.

After we found a $t$-valid polynomial $\Lambda(x)$, its $t$ roots $\alpha^{-j_1}, \alpha^{-j_2}, \ldots, \alpha^{-j_t}$ indicate the positions $j_1, j_2, \ldots, j_t$ of the erroneous symbols in the received word $y(x)$. Error evaluation, i.e., determining the values of the error symbols can then be performed with respect to the original Reed–Solomon code $\mathcal{C}$ using standard methods like the Recursive Extension [10], or by adopting the Forney Algorithm [9].

Based on syndrome extension, Reed–Solomon decoding beyond half the minimum distance can be performed by the following steps: For a corrupted word $y(x)$ of a Reed–Solomon code $\mathcal{C} = \mathcal{RS}(q; n, k)$ of rate $R \lesssim 1/3$, we first determine the largest integer $l$, which satisfies (11). Then, we create a matrix $\boldsymbol{Y}$ accordant to (6) by raising the received symbols $y_0, y_1, \ldots, y_{n-1}$ to the $i$th power for $i = 1, \ldots, l$. We consider $\boldsymbol{Y}$ as received word of a heterogeneous IRS code and calculate a syndrome sequence for any row pursuant to (7). Then, we use Algorithm 1 to synthesize a pair $t, \Lambda(x)$. If $\Lambda(x)$ is a $t$-valid polynomial, then we find the roots $\alpha^{-j_1}, \alpha^{-j_2}, \ldots, \alpha^{-j_t}$, and evaluate the errors at the positions $j_1, j_2, \ldots, j_t$ by the Forney algorithm, by Syndrome Extension, or by any other standard method for error evaluation. The complete decoding procedure is summarized by Algorithm 2. Generally, depending on the received word $y(x)$, Algorithm 2 may yield the following results:

1) The algorithm may yield the transmitted codeword, i.e, $\hat{c}(x) \equiv c(x)$.

```
input: received word y(x) = r_0 + y_1 x + ··· + y_{n-1} x^{n-1}
determine largest integer l satisfying (11)
for each i from 1 to l do
    compute y^⟨i⟩(x) = y_0^i + y_1^i x + ··· + y_{n-1}^i x^{n-1}
create matrix Y from y^⟨i⟩(x), i = 1, ..., l
use DFT to calculate syndromes 𝒮^[1], ..., 𝒮^[l]
compute t, Λ(x) by Algorithm 1
if t ≤ t_max^{[l]} and Λ(x) is t-valid then
    evaluate errors, and calculate e(x)
    calculate ĉ(x) = y(x) - e(x)
else
    decoding failure
output: ĉ(x) ∈ 𝒞 or decoding failure
```

Algorithm 2.  Decoder Based on Syndrome Extension

2) The algorithm may decode a wrong codeword, i.e., $\hat{c}(x) \not\equiv c(x)$.
3) The algorithm may not yield a codeword at all, i.e, it may yield a decoding failure.

These tree events occur with a certain probability. In the following, the probability or a correct decision is denoted by $P_c$, the probability for a decoding error is denoted by $P_e$, and the failure probability by $P_f$. Hence, the probability $P_w$ for an incorrect decoding result is obtained by $P_w = 1 - P_c = P_e + P_f$.

Algorithm 2 possesses an interesting property, which is called *Maximum Likelihood certificate property*. The term ML certificate property was introduced in [26], where it is shown that a decoding algorithm based on linear programming exhibits this property. Formally, we define the ML certificate property as follows:

**Definition 4 (ML Certificate Property)**
*Consider a code $\mathcal{C}$, and assume that the word $\boldsymbol{y}$ is received when a codeword $\boldsymbol{c} \in \mathcal{C}$ is transmitted over a memoryless noisy channel. Moreover, consider a decoding algorithm which either decides on a unique codeword, or yields a decoding failure. We say that the decoding algorithm exhibits the ML certificate property, if it behaves in such a way that whenever it yields a codeword $\hat{\boldsymbol{c}} \in \mathcal{C}$, this codeword fulfills*

$$\text{dist}(\boldsymbol{y}, \hat{\boldsymbol{c}}) = \min_{\boldsymbol{c} \in \mathcal{C}} \{\text{dist}(\boldsymbol{y}, \boldsymbol{c})\}.$$

In [20], a IRS decoding algorithm based on multi–sequence shift–register synthesis is proposed. Moreover, it is shown that this algorithm possesses the ML certificate property. After creating the matrix $\boldsymbol{Y}$ from the received word $y(x)$, Algorithm 2 considers $\boldsymbol{Y}$ as received matrix of an IRS code and processes it exactly like the IRS decoding algorithm proposed in [20]. Thus, the correctness of the following theorem directly follows from the results proved in [20].

**Theorem 1** *Algorithm 2 possesses the ML certificate property.*

## VI. Error and Failure Probability

The syndrome extension procedure described above increases the maximum error correcting radius by increasing the number of linear equations for calculating the error locator polynomial. In this case, (8) *can* have a *unique* solution for a larger number of unknowns $t$. However, it really only has a *unique* solution, if the additional equations obtained from the auxiliary syndromes are linearly independent, i.e., if $\mathrm{rank}(\boldsymbol{S}) = t$. Otherwise, a decoding failure is obtained. Hence, syndrome extension effectively increases the decoding performance only, if the failure probability $P_f$ is fairly small. Consequently, to show that the proposed decoding method effectively works, we have to analyze its failure and error probability.

### A. Error Probability of Syndrome Extension

It immediately follows from Theorem 1, that the error probability $P_e$ is not larger than the error probability of an ML decoder. Moreover, we are able to use the following result from [20]:

**Theorem 2 (Error Probability)** *Let $\mathcal{C}(q; n, k, d)$ be a linear block code of length $n$, dimension $k$, and minimum distance $d$ over the field $\mathbb{F}_q$, decoded by a BD decoder which exhibits the ML certificate property. Assume that the decoding radius of this decoder is $t_{\max}$, and that it decodes a received word, which is corrupted by $t$ errors. Then, the probability for a decoding error is overbounded by*

$$P_e(t) \le \overline{P}_e(t) = \frac{\sum_{w=d}^{t+t_{\max}} A_w \sum_{\rho=0}^{\min\{t,t_{\max}\}} \cdot U_q(t,w,\rho)}{\binom{n}{t}(q-1)^t}, \tag{15}$$

*where $A_w$ is the weight distribution of the code, that is, $A_w$ is the number of codewords of weight $w$ in $\mathcal{C}$, and $U_q(r_2, r_1, \rho)$ is calculated by*

$$U_q(r_2, r_1, \rho) = \sum_{i=\lceil \frac{r_1+r_2-\rho}{2} \rceil}^{r_1+r_2-\rho} \binom{r_1}{i}\binom{i}{\rho-(r_1+r_2)+2i} \cdot \\ \cdot \binom{N-r_1}{r_2-i} \cdot (q-2)^{\rho-(r_1+r_2)+2i}(q-1)^{r_2-i}. \tag{16}$$

Since Algorithm 2 possesses the ML certificate property, Theorem 2 can directly be applied for overbounding the error probability $P_e$.

### B. Failure Probability

Even though $\boldsymbol{Y}$ is considered as a received word of a heterogeneous IRS code, we cannot apply the upper bound on the failure probability derived in [20]. The reason for this lies in the fact, that for deriving the bound in [20], it is assumed that the erroneous columns in the received matrices are distributed uniformly over all non-zero vectors. Obviously this is not true for a heterogeneous IRS code obtained by syndrome extension, since all symbols in an erroneous column $j$ are just powers of the error symbol $e_j$. Consequently, the analysis of the failure probability of syndrome extension decoding is substantially more involved than the analysis of the failure probability for ordinary IRS decoding. However, for fields of characteristic 2 and for $l = 2$, i.e., for Reed–Solomon codes of rate $1/6 \lesssim R \lesssim 1/3$ defined over the extension field $\mathbb{F}_{2^m}$, we are nevertheless able to derive a bound similar to the bound for ordinary IRS codes derived in [20]. This allows for an analytical evaluation of the decoding performance of Algorithm 2 for the most interesting low–rate Reed–Solomon codes. Later, we complement this analysis by experimental results obtained from Monte–Carlo simulations, which show that Algorithm 2 also works quite well, when it is applied to Reed–Solomon codes of rate smaller than $1/6$, i.e., for $l > 2$.

For a Reed–Solomon code $\mathcal{C}$ over $\mathbb{F}_{2^m}$ with a rate in the range $1/6 \lesssim R \lesssim 1/3$, for which the largest integer satisfying (11) is $l = 2$, Algorithm 2 creates a matrix $\boldsymbol{Y}$ with two rows by squaring the coefficients of the received word $y(x)$. Since the code $\mathcal{C}$ is defined over a field of characteristic 2, we have $r_i^2 = (a_i + e_i)^2 = a_i^2 + e_i^2$, which means that $\boldsymbol{Y}$ can be written as

$$\boldsymbol{Y} = \begin{pmatrix} y_0 & y_1 & \cdots & y_{n-1} \\ y_0^2 & y_1^2 & \cdots & y_{n-1}^2 \end{pmatrix} = \\ = \underbrace{\begin{pmatrix} c_0 & c_1 & \cdots & c_{n-1} \\ c_0^2 & c_1^2 & \cdots & c_{n-1}^2 \end{pmatrix}}_{\boldsymbol{C}} + \underbrace{\begin{pmatrix} e_0 & e_1 & \cdots & e_{n-1} \\ e_0^2 & e_1^2 & \cdots & e_{n-1}^2 \end{pmatrix}}_{\boldsymbol{E}}. \tag{17}$$

For a matrix $\boldsymbol{Y}$ in this form, we are able to derive an upper bound on the failure probability of Algorithm 2. To do this, we overbound the number of cases, in which the linear system of equations (8) has multiple solutions. Pursuant to the following lemma, which is a variant of Lemma 5 from [20], this yields an upper bound on the failure probability of Algorithm 2:

**Lemma 3** *Consider a codeword $c(x)$ of a Reed–Solomon code $\mathcal{C}$, for which (11) is satisfied with $l > 1$. Assume that this word is corrupted by an error of weight $t \le t_{\max}$ and that Algorithm 2 yields a decoding failure. Then, the linear system of equations (8) with $t$ unknowns has multiple solutions.*

*Proof:* Assume that a codeword $c(x) \in \mathcal{C}$ has been transmitted, and that the channel adds an error word $e(x)$ with $t$ non-zero elements at the positions $j_1, j_2, \ldots, j_t$. Then Algorithm 2 creates a matrix $\boldsymbol{Y}$ with $t$ erroneous columns. Algorithm 2 only yields a decoding failure, if Algorithm 1 computes a pair $t', \Lambda'(x)$, such that $\Lambda'(x)$ is not $t'$-valid.

First we show, that $t' \le t$. We know, that the polynomial $\Lambda(x) = \prod_{i=1}^{t}(1 - \alpha^{j_i}x)$ is a $t$-valid polynomial, which represents a solution of the set of linear recursions (14). Since Algorithm 1 always yields the smallest length shift–register which is a solution of (14), we know that $t' \le t$.

Second, $\Lambda'(x)$ is a solution of the set of linear recursions (14) of length $t'$, and hence also for the system of equations (8) with $t'$ unknowns. However, if $\Lambda'(x)$ is a solution for the set of linear recursions (14) of length $t'$, it is also a solution for length $t \ge t'$, since if $\Lambda'(x)$ fulfills (14) for $i \ge t'$, it obviously also fulfills (14) for $i \ge t \ge t'$. Thus, $\Lambda'(x)$ is not only a solution of the system of equations (8) with $t'$ unknowns and the set of linear recursions (14) of length $t'$, but also for the system (8) with $t$ unknowns, and the set of recursions (14) of length $t$. Consequently, the system of equations (8) with $t$ unknowns has at least the two solutions $\Lambda(x)$ and $\Lambda'(x)$. This proves the statement of Lemma 3. ∎

Note, that from Lemma 3 we only know that (8) always has multiple solutions if Algorithm 2 fails. Contrariwise this does not mean that Algorithm 2 *always* fails if (8) has multiple solutions. Nevertheless, by counting the cases in which (8) multiple solutions, we obtain an upper bound on the probability $P_f$ that Algorithm 2 fails.

In the next step, we consider the following lemma, which allows us to utilize the structure of a matrix $Y$ pursuant to (17):

**Lemma 4** *Let $(v, e)$ be a pair of two non-zero field elements. Then, the set*
$$\mathcal{V} = \left\{ \begin{pmatrix} ve \\ ve^2 \end{pmatrix} : v, e \in \mathbb{F}_q \setminus \{0\} \right\}$$
*forms the set of all full weight vectors of length 2, i.e.,*
$$\mathcal{V} \equiv \left\{ \boldsymbol{v} \in \mathbb{F}_q^2 : \mathrm{wt}\,(\boldsymbol{v}) = 2 \right\} = [\mathbb{F}_q \setminus \{0\}]^2 \;.$$

*Proof:* Since there exist $(q-1)^2$ possibilities to select the pairs $(v, e)$, it is sufficient to show that all $(q-1)^2$ vectors $\boldsymbol{v} \in \mathcal{V}$ are pairwise different. Indeed, by considering two vectors $\boldsymbol{v}, \tilde{\boldsymbol{v}} \in \mathcal{V}$, and assuming that $\boldsymbol{v} = \tilde{\boldsymbol{v}}$, we get
$$ve = \tilde{v}\tilde{e}, \quad (18)$$
$$ve^2 = \tilde{v}\tilde{e}^2 \;. \quad (19)$$
Dividing (19) by (18) yields $e = \tilde{e}$, and by inserting this into (18) we obtain $v = \tilde{v}$. Hence, two different pairs $(v, e) \neq (\tilde{e}, \tilde{v})$ always yield two different vectors $\boldsymbol{v}, \tilde{\boldsymbol{v}}$. ∎

Next, we consider the following lemma, which is a variant of Lemma 6 from [20]:

**Lemma 5** *Let $\mathcal{W}^{(1)}, \mathcal{W}^{(2)}$ be two q-ary linear codes of length $\omega$, and let the dimension of $\mathcal{W}^{(1)}$ be $\omega - \varrho^{(1)}$, and the dimension of $\mathcal{W}^{(2)}$ be $\omega - \varrho^{(2)}$. Further, let*
$$\boldsymbol{W} = \begin{pmatrix} \boldsymbol{w}^{(1)} \\ \boldsymbol{w}^{(2)} \end{pmatrix} = (\boldsymbol{w}_0, \ldots, \boldsymbol{w}_{\omega-1})$$
*be a $2 \times \omega$ matrix such that $\mathrm{wt}\,(\boldsymbol{w}_j) = 2, \forall j = 0, \ldots, \omega-1$, i.e., $\boldsymbol{W}$ does not contain any zero element. Moreover, assume that all columns of $\boldsymbol{W}$ are uniformly distributed over all full weight vectors of length 2. Then, the probability $P_\omega$ that*
$$\boldsymbol{w}^{(1)} \in \mathcal{W}^{(1)}, \; \boldsymbol{w}^{(2)} \in \mathcal{W}^{(2)} \quad (20)$$
*is overbounded by*
$$P_\omega \leq \frac{q^{2\omega}}{(q-1)^{2\omega}} \cdot q^{-\left(\varrho^{(1)} + \varrho^{(2)}\right)} \;. \quad (21)$$

*Proof:* Let $\mathcal{L}$ be the set of all $2 \times \omega$ matrices whose rows fulfill (20). Further, let $\mathcal{S}_\omega$ be the set of all $2 \times \omega$ matrices with elements from $\mathbb{F}_q$, and let the subset $\mathcal{S}'_\omega \subset \mathcal{S}_\omega$ be the set of matrices without any non-zero elements. Then, the probability $P_\omega$ for a matrix $\boldsymbol{W}$ without zero elements to fulfill (20) is overbounded by
$$P_\omega = \frac{|\mathcal{L} \cap \mathcal{S}'_\omega|}{|\mathcal{S}'_\omega|} \leq \frac{|\mathcal{L}|}{|\mathcal{S}'_\omega|} \;.$$
The cardinality $|\mathcal{L}|$ is calculated by by
$$|\mathcal{L}| = |\mathcal{W}^{(1)}| \cdot |\mathcal{W}^{(1)}| = q^{2\omega - \left(\varrho^{(1)} + \varrho^{(2)}\right)}\;,$$
and $|\mathcal{S}'_\omega|$ is obtained by
$$|\mathcal{S}'_\omega| = (q-1)^{2\omega} \;.$$
Dividing $|\mathcal{L}|$ by $|\mathcal{S}'_\omega|$ yields the right hand side of (21), which proves Lemma 5. ∎

By utilizing Lemma 3, Lemma 4, and Lemma 5, we are now able to overbound the failure probability for codes over fields of characteristic two and for $l = 2$ by the following theorem:

**Theorem 3 (Failure Probability)** *Let $\mathcal{C}$ be a Reed–Solomon code over $\mathbb{F}_{2^m}$, for which $l = 2$ is the largest integer satisfying (11). Assume that a codeword $a(x) \in \mathcal{C}$ is corrupted by an error word $e(x)$ with $t$ non-zero coefficients, and that the resulting word $r(x) = a(x) + e(x)$ is decoded by Algorithm 2. Then, the probability for a decoding failure is overbounded by*
$$P_f(t) \leq \overline{P}_f(t) = \underbrace{\left( \frac{q}{q-1} + \frac{1}{q} \right)^t}_{\gamma} \cdot \frac{q^{-3 \cdot (t_{\max} - t)}}{q-1}\;, \quad (22)$$
*where $t_{\max} = t_{\max}^{[2]}$ is calculated by Equation (9).*

*Proof:* According to Lemma 3, the failure probability of Algorithm 2 can be overbounded by investigating the linear system of equations (8). More precisely, we are interested in the cases in which (8) has multiple solutions, i.e., the cases in which $\mathrm{rank}\,(\boldsymbol{S}) < t$. We have such a case, if there exist a vector $\boldsymbol{u} \neq \boldsymbol{0}$ such that $\boldsymbol{S} \cdot \boldsymbol{u} = \boldsymbol{0}$. Equivalently, we can say that (8) cannot have a unique solution, if
$$\exists\, \boldsymbol{u} \neq \boldsymbol{0} \,:\, \boldsymbol{S}^{[i]} \cdot \boldsymbol{u} = \boldsymbol{0} \;\; \forall\, i = 1, 2 \;. \quad (23)$$

Since both syndrome matrices $\boldsymbol{S}^{[1]}$ and $\boldsymbol{S}^{[2]}$ are determined by the same error word $\boldsymbol{e} = (e_0, e_1, \ldots, e_{n-1})$, we are able to express the failure probability $P_f(t)$ by
$$P_f(t) = \frac{|\{\boldsymbol{e} \,:\, \mathrm{wt}\,(\boldsymbol{e}) = t \text{ and (23) holds}\}|}{|\{\boldsymbol{e} \,:\, \mathrm{wt}\,(\boldsymbol{e} = t)\}|} \;.$$
Now, we fix $t$ positions $j_1, j_2, \ldots, j_t$, and define the ensemble $\mathcal{E}_t(j_1, \ldots, j_t)$ of equiprobable vectors $\boldsymbol{e}$ of weight $\mathrm{wt}\,(\boldsymbol{e}) = t$ that may take all possible non-zero values at the indices $j_1, j_2, \ldots, j_t$, and are zero in all remaining coordinates. The cardinality of this ensemble is calculated by $|\mathcal{E}_t(j_1, \ldots, j_t)| = (q-1)^t$. Moreover, we define the probability $P_f(j_1, \ldots, j_t)$ that (8) is satisfied for vectors from the ensemble $\mathcal{E}_t(j_1, \ldots, j_t)$ by
$$P_f(j_1, \ldots, j_t) = \frac{\overbrace{|\{\boldsymbol{e} \in \mathcal{E}_t(j_1, j_2, \ldots, j_t) \,:\, (23) \text{ holds}\}|}^{N_b}}{|\mathcal{E}_t(j_1, \ldots, j_t)|} \;.$$
(24)

For overbounding $P_f(j_1, \ldots, j_t)$, let the number of rows in $\boldsymbol{S}^{[1]}$ be denoted by $\varrho^{[1]} = n - k - t$, and the number of rows in $\boldsymbol{S}^{[2]}$ be denoted by $\varrho^{[1]} = n - 2k - t + 1$. It is known (cf. e.g. [10]) that a syndrome matrices $\boldsymbol{S}^{[i]}$, $i = 1, 2$, can be decomposed into

$$\boldsymbol{S}^{[i]} = \boldsymbol{H}^{[i]} \cdot \boldsymbol{F}^{[i]} \cdot \boldsymbol{D} \cdot \boldsymbol{V_t},$$

where

$$\boldsymbol{V_t} = \begin{pmatrix} 1 & \alpha^{j_1} & \alpha^{2j_1} & \ldots & \alpha^{(t-1)j_1} \\ 1 & \alpha^{j_2} & \alpha^{2j_2} & \ldots & \alpha^{(t-1)j_2} \\ \vdots & \vdots & & \vdots & \\ 1 & \alpha^{j_t} & \alpha^{2j_t} & \ldots & \alpha^{(t-1)j_t} \end{pmatrix}$$

is a $t \times t$ *Vandermonde matrix*, the matrices

$$\boldsymbol{D} = \operatorname{diag}\left(\alpha^{j_1}, \alpha^{j_2}, \ldots, \alpha^{j_t}\right)$$

and

$$\boldsymbol{F}^{[i]} = \operatorname{diag}\left(e_{j_1}^i, e_{j_2}^i, \ldots, e_{j_t}^i\right)$$

are $t \times t$ diagonal matrices, and the matrix

$$\boldsymbol{H}^{[i]} = \begin{pmatrix} 1 & 1 & \ldots & 1 \\ \alpha^{j_1} & \alpha^{j_2} & \ldots & \alpha^{j_t} \\ \alpha^{2j_1} & \alpha^{2j_2} & \ldots & \alpha^{2j_t} \\ \vdots & \vdots & & \vdots \\ \alpha^{(\varrho^{[i]}-1)j_1} & \alpha^{(\varrho^{[i]}-1)j_2} & \ldots & \alpha^{(\varrho^{[i]}-1)j_t} \end{pmatrix}$$

is a $\varrho^{[i]} \times t$ matrix consisting of $\varrho^{[i]}$ rows of a transposed Vandermonde matrix. In other words, $\boldsymbol{H}^{[i]}$ can be considered as parity-check matrix of a (shortened) Reed–Solomon code of length $t$ and dimension $t - \varrho^{[i]}$, which we denote by $\mathcal{W}^{(i)}$. Since the matrices $\boldsymbol{V_t}$ and $\boldsymbol{D}$ both have full rank, the product $\boldsymbol{v} = \boldsymbol{D}\boldsymbol{V_t} \cdot \boldsymbol{u}$ defines a bijective linear map $\boldsymbol{u} \mapsto \boldsymbol{v}$. Consequently, the statement

$$\exists\, \boldsymbol{v} \neq \boldsymbol{0} \,:\, \boldsymbol{H}^{[i]} \boldsymbol{F}^{[i]} \cdot \boldsymbol{v} = \boldsymbol{0} \;\; \forall\, i = 1, 2 \quad (25)$$

is equivalent to Equation (23). Due to this equivalence, we are able to calculate $N_b$ by

$$N_b = |\{\boldsymbol{e} \in \mathcal{E}_t(j_1, \ldots, j_t) : (23) \text{ holds}\}| =$$
$$= |\{\boldsymbol{e} \in \mathcal{E}_t(j_1, \ldots, j_t) : (25) \text{ holds}\}|.$$

However, instead of directly counting the error vectors $\boldsymbol{e}$ satisfying (25), we consider all pairs $(\boldsymbol{v}, \boldsymbol{e})$ of vectors, which satisfy

$$\boldsymbol{H}^{[i]} \boldsymbol{F}^{[i]} \cdot \boldsymbol{v} = \boldsymbol{0} \;\; \forall\, i = 1, 2. \quad (26)$$

In this way, we are able to overbound $N_b$ by

$$N_b \leq \left|\{(\boldsymbol{v}, \boldsymbol{e}) : \boldsymbol{e} \in \mathcal{E}_t(j_1, \ldots, j_t), \boldsymbol{v} \in \mathbb{F}_q^t \setminus \{\boldsymbol{0}\}, (26) \text{ holds}\}\right| \quad (27)$$

This is a Union Bound, where error vectors $\boldsymbol{e}$ satisfying (25) may be counted several times with different vectors $\boldsymbol{v}$.

To improve (27), we take into account that if (26) holds for a pair $(\boldsymbol{v}, \boldsymbol{e})$, then it also holds for all pairs $(\alpha \boldsymbol{v}, \boldsymbol{e})$, $\alpha \in \mathbb{F}_q \setminus \{0\}$. Hence, every vector $\boldsymbol{e}$ is counted in (27) at least $q - 1$ times, which means that we are able to improve (27) by the factor $q - 1$. In this way, we obtain

$$N_b \leq \tfrac{1}{q-1} \sum_{\boldsymbol{v} \in \mathbb{F}_q^t \setminus \{\boldsymbol{0}\}} \underbrace{|\{(\boldsymbol{v}, \boldsymbol{e}) : \boldsymbol{e} \in \mathcal{E}_t(j_1, \ldots, j_t), (26) \text{ holds}\}|}_{M(\boldsymbol{v})}.$$

Partitioning $\{\boldsymbol{v} \in \mathbb{F}_q^t \setminus \{\boldsymbol{0}\}\}$ into the sets $\mathcal{V}_\omega(i_1, \ldots, i_\omega)$ of vectors $\boldsymbol{v}$ with $\omega$ non-zero elements at the indices $i_1, i_2, \ldots, i_\omega$ yields

$$N_b \leq \tfrac{1}{q-1} \sum_{\omega=1}^{t} \sum_{(i_1, \ldots, i_\omega)} \sum_{\boldsymbol{v} \in \mathcal{V}_\omega(i_1, \ldots, i_\omega)} M(\boldsymbol{v}), \quad (28)$$

and by inserting (28) into (24), we obtain

$$P_f(j_1, \ldots, j_t) \leq$$
$$\leq \frac{1}{q-1} \sum_{\omega=1}^{t} \sum_{(i_1, \ldots, i_\omega)} \sum_{\boldsymbol{v} \in \mathcal{V}_\omega(i_1, \ldots, i_\omega)} \frac{M(\boldsymbol{v})}{|\mathcal{E}_t(j_1, \ldots, j_t)|}. \quad (29)$$

Now, we reconsider the fact that $\boldsymbol{H}^{[i]}$ is a parity–check matrix of the code $\mathcal{W}^{[i]}$, and denote by $\boldsymbol{w}^{[i]} = \left(\boldsymbol{F}^{[i]} \cdot \boldsymbol{v}\right)^\mathsf{T}$ the vector obtained by multiplying $\boldsymbol{v}$ and $\boldsymbol{F}^{[i]}$. This enables us to state another condition, which is equivalent to (26):

$$\boldsymbol{w}^{[i]} \in \mathcal{W}^{[i]} \;\; \forall\, i = 1, 2. \quad (30)$$

Consequently, instead of a pair $(\boldsymbol{v}, \boldsymbol{e})$, we equivalently may consider the matrix

$$\boldsymbol{W} = \begin{pmatrix} \boldsymbol{w}^{[1]} \\ \boldsymbol{w}^{[2]} \end{pmatrix} = \begin{pmatrix} v_1 e_{j_1} & v_2 e_{j_2} & \ldots & v_t e_{j_t} \\ v_1 e_{j_1}^2 & v_2 e_{j_2}^2 & \ldots & v_t e_{j_t}^2 \end{pmatrix},$$

and write $\boldsymbol{H}^{[i]} \boldsymbol{W}^\mathsf{T} = \boldsymbol{0}$ instead of (26). For a vector $\boldsymbol{v}$ with non-zero elements at the indices $i_1, i_2, \ldots, i_\omega$, the corresponding matrix $\boldsymbol{W}$ possesses exactly $\omega$ non-zero columns, from which we create a $2 \times \omega$ matrix $\boldsymbol{W}_\omega$. Moreover, by creating the $2 \times \omega$ matrix $\boldsymbol{H}_\omega^{[i]}$ from the columns $i_1, i_2, \ldots, i_\omega$ of $\boldsymbol{H}^{[i]}$, the statements $\boldsymbol{H}^{[i]} \boldsymbol{W}^\mathsf{T} = \boldsymbol{0}$ and

$$\boldsymbol{H}_\omega^{[i]} \boldsymbol{W}_\omega^\mathsf{T} = \boldsymbol{0} \quad (31)$$

are equivalent. Now, we show that the total number $|\boldsymbol{W}_\omega|$ of matrices satisfying (31), can be calculated by

$$P_\omega \cdot |\boldsymbol{W}_\omega| = P_\omega \cdot |\mathcal{V}_\omega(i_1, \ldots, i_\omega)| \cdot |\mathcal{E}_t(j_1, \ldots, j_t)|. \quad (32)$$

To do this, we consider the set of all matrices satisfying (31), which are obtainable from all vectors $\boldsymbol{v} \in \mathcal{V}_\omega(i_1, \ldots, i_\omega)$ and $\boldsymbol{e} \in \mathcal{E}_t(j_1, \ldots, j_t)$. We observe that the elements $v_j$ and $e_{j_i}$ in these matrices may independently take all values from $\mathbb{F}_q \setminus \{0\}$. Hence, by applying Lemma 4, we conclude that the columns of all matrices $\boldsymbol{W}_\omega$ are independent random vectors from $[\mathbb{F}_q \setminus \{0\}]^2$. This in turn enables us to apply Lemma 5 for overbounding the probability $P_\omega$ for a matrix $\boldsymbol{W}_\omega$ to satisfy

(31). Finally, by combining (29) and (32) we obtain

$$P_f(j_1,\ldots,j_t) \le \frac{1}{q-1} \sum_{\omega=1}^{t} \sum_{(i_1,\ldots,i_\omega)} (q-1)^\omega P_\omega =$$
$$= \frac{1}{q-1} \sum_{\omega=1}^{t} \binom{t}{\omega} \frac{q^{2\omega}}{(q-1)^\omega} \cdot q^{-(\varrho^{(1)}+\varrho^{(2)})} =$$
$$= \frac{q^{-(\varrho^{(1)}+\varrho^{(2)})}}{q-1} \sum_{\omega=1}^{t} \binom{t}{\omega} \left(\frac{q^2}{q-1}\right)^\omega \le$$
$$\le \frac{q^{-(\varrho^{(1)}+\varrho^{(2)})}}{q-1} \sum_{\omega=0}^{t} \binom{t}{\omega} \left(\frac{q^2}{q-1}\right)^\omega =$$
$$= \left(\frac{q^2}{q-1}+1\right)^t \frac{q^{-(2n-3k-2t+1)}}{q-1} =$$
$$= \left(\frac{q}{q-1}+\frac{1}{q}\right)^t \cdot \frac{q^{-3 \cdot (t_{\max}-t)}}{q-1} . \quad (33)$$

Since the right side of (33) does not depend on the selection of $j_1, j_2, \ldots, j_t$, but only on weight $t$ of the error vector $e$, it also provides us with an upper bound on $P_f(t)$. This proves Theorem 3. ∎

Since the factor $\gamma$ is close to one, the failure probability $P_f(t_{\max})$ is in the order of $1/(q-1)$ for the maximum error correcting radius $t_{\max}$. Moreover we see, that the failure probability decreases exponentially with decreasing $t$. Hence, (22) behaves rather similar to the upper bound on the failure probability for ordinary IRS codes. The only difference is, that for an IRS code obtained by syndrome extension, the factor $\gamma$ is slightly larger.

### C. Probability $P_w$ in the QSC

If we use Algorithm 2 for decoding a Reed–Solomon codeword which is corrupted by $t \le \tau = \lfloor \frac{n-k}{2} \rfloor$ errors, the decoding result is always unique and we are always able to correct the errors. Hence, the probability for an incorrect decoding result is $P_w = 0$ in this case. If the number of errors is in the range $\tau < t \le t_{\max}^{[l]}$, Algorithm 2 yields a decoding failure with probability $P_f > 0$, and a decoding error with probability $P_e > 0$. However, since the probability for an incorrect decoding result $P_w = P_e + P_f$ is generally much smaller than one, we are still able to correct the errors with the probability $P_c > 0$. If $t > t_{\max}^{[l]}$, Algorithm 2 either yields a decoding failure, or an erroneous decoding result. Hence, we have the probability $P_w = P_e + P_f = 1$ in this case. Based on these facts, the probability $P_w$ for an incorrect decoding result in the QSC with crossover probability $p$ is calculated in the usual way by

$$P_w = \sum_{t=\tau+1}^{n} \binom{n}{t} \cdot (P_e(t) + P_f(t)) \cdot p^t \cdot (1-p)^{n-t} .$$

Clearly, for Reed–Solomon codes with rate $1/6 \lesssim R \lesssim 1/3$ we are able to overbound $P_w$ by replacing $P_e(t)$ and $P_f(t)$ by their upper bounds (15) and (22) respectively. More precisely, we replace $P_w(t)$ by

$$\overline{P}_w(t) \triangleq \min\left\{\overline{P}_e(t) + \overline{P}_f(t), 1\right\} , \quad (34)$$

where $\overline{P}_e(t)$ is computed by means of Theorem 2 and $\overline{P}_f(t)$ is calculated by means of Theorem 3. In this way, we obtain the following corollary:

**Corollary 4** *Let $\mathcal{C}$ be a Reed–Solomon code over $\mathbb{F}_{2^m}$ for which $l = 2$ is the largest integer satisfying (11). Assume that the codewords of this codes are transmitted over a QSC with crossover probability $p$, and that the received words are decoded by Algorithm 2. Then, the probability $P_w^{l=2}$ to obtain an incorrect decoding result is overbounded by*

$$P_w^{l=2} \le \overline{P}_w^{l=2} = \sum_{t=\tau+1}^{n} \binom{n}{t} \cdot \overline{P}_w(t) \cdot p^t \cdot (1-p)^{n-t} , \quad (35)$$

*where $\overline{P}_w(t)$ is given by (34).*

### D. Error Probability versus Failure Probability

Comparing the upper bound (15) on the error probability $P_e$ and the upper bound (22) on the failure probability $P_f$ yields that the error probability is generally much smaller than the failure probability. To demonstrate this, we consider the Reed–Solomon code $\mathcal{RS}\left(2^8, 255, 63\right)$. By classical BMD decoding we are able to correctly decode all error patterns with $t \le 96$ non-zero symbols. Moreover, Algorithm 2 is able to decode error patterns in up to $t_{\max} = 107$. Calculating the upper bounds (15) and (22) for the interval $97 \le t \le 107$ yields the results depicted in Fig. 3. We observe that the error probability $P_e$ is smaller by more than 100 orders of magnitude than the failure probability $P_f$. This means, that from a practical point of view the error probability $P_e$ is absolutely negligible for the decoding performance of Algorithm 2. The probability $P_w$ for an incorrect decoding result is absolutely dominated by the failure probability $P_f$. Consequently, Theorem 3 should be considered as one of the most important results in this paper, since it theoretically proves the effectiveness of syndrome extension decoding.

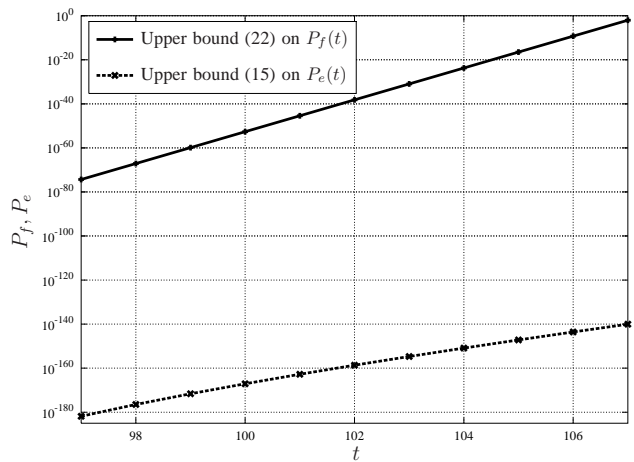

Fig. 3. Upper bounds on the failure and error probability for a Reed–Solomon code $\mathcal{RS}\left(2^8; 255, 63\right)$ decoded by Algorithm 2.

## E. Experimental Results

To verify the tightness of the bound stated by Theorem 3, we would like to compare it with the actual failure probability obtained by Algorithm 2. In principle, we are able to obtain the actual failure probability by Monte–Carlo simulations. However, since $\overline{P}_f$ decreases with the field cardinality $|\mathbb{F}_q|$, reliable results can only be obtained for Reed–Solomon codes over fields with small cardinalities. Therefore we considered Reed–Solomon codes over the field $\mathbb{F}_{2^5}$ for our experimental investigations.

Pursuant to the equations (10) and (13), the Reed–Solomon code $\mathcal{RS}\left(2^5, 31, 6\right)$ can be decoded by Algorithm 2 using $l = 2$ syndromes up to the maximum decoding radius $t_{\max}^{[2]} = 15$. In contrast to this, a classical BMD decoder is able to correct up to 12 errors. Thus, we performed experiments in which we created $N = 10^8$ random error vectors of weight $t$ for each $t$ in the interval $12 < t \leq 15$, and counted the number $N_f$ of decoding failures. The results of these experiments are presented in Table I.

TABLE I

EXPERIMENTAL RESULTS ON THE FAILURE PROBABILITY $P_f(t)$ FOR THE REED–SOLOMON CODE $\mathcal{RS}(31, 6)$

| $\mathcal{RS}\left(2^5; 31, 6\right)$ | | | |
| --- | --- | --- | --- |
| $l$ | 2 | | |
| $t_{\max}^{[2]}$ | 15 | | |
| $t$ | 13 | 14 | 15 |
| $N$ | $10^8$ | $10^8$ | $10^8$ |
| $N_f$ | 0 | 88 | 3025500 |
| $P_f$ | $\lesssim 10^{-8}$ | $\approx 9 \cdot 10^{-7}$ | $\approx 3 \cdot 10^{-2}$ |
| $\overline{P}_f$ (22) | $6.7 \cdot 10^{-11}$ | $2.3 \cdot 10^{-6}$ | $8.1 \cdot 10^{-2}$ |

From Table I we observe, that Theorem 3 gives a quite good estimate for the actual behavior of the failure probability for Reed–Solomon codes over fields of characteristic 2 decoded by Algorithm 2 with $l = 2$, i.e., for Reed–Solomon codes of rate $1/6 \lesssim R \lesssim 1/3$. We see that $P_f(t)$ actually decreases exponentially with decreasing $t$ like predicted by Theorem 3.

To investigate whether we have a similar behavior for $l > 3$, we consider the Reed–Solomon code $\mathcal{RS}\left(2^5; 31, 4\right)$, which can be decoded by a BMD decoder up to the radius 14, and by Algorithm 2 with $l = 3$ up to the radius $t_{\max}^{[3]} = 18$. We again performed $N = 10^8$ random experiments for each $t$ in the range $14 < t \leq 18$. Table II shows the results of these experiments.

TABLE II

EXPERIMENTAL RESULTS ON THE FAILURE PROBABILITY $P_f(t)$ FOR THE REED–SOLOMON CODE $\mathcal{RS}(31, 4)$

| $\mathcal{RS}\left(2^5; 31, 4\right)$ | | | | |
| --- | --- | --- | --- | --- |
| $l$ | 3 | | | |
| $t_{\max}^{[3]}$ | 18 | | | |
| $t$ | 15 | 16 | 17 | 18 |
| $N$ | $10^8$ | $10^8$ | $10^8$ | $10^8$ |
| $N_f$ | 0 | 0 | 37 | 3121501 |
| $P_f$ | $\lesssim 10^{-8}$ | $\lesssim 10^{-8}$ | $\approx 4 \cdot 10^{-7}$ | $\approx 3 \cdot 10^{-2}$ |

Form Table II we see the basic behavior of $P_f(t)$ for $l > 2$ is similar to the case $l = 2$, even if we are not able to derive an upper bound for this case. Hence, from a practical point of view, we can say that our syndrome extension procedure can effectively be applied for all Reed–Solomon codes of rate $R \lesssim 1/3$.

## VII. DECODING PERFORMANCE

As already mentioned, the threshold rates and the error correcting radii of Algorithm 2 are similar to the classical Sudan algorithm [3]. However, since the decoding strategy in the case of an ambiguous decoding result is different, the question is how our syndrome extension approach actually performs in comparison to the Sudan algorithm. To study this, and to give a vivid expression of what can be gained by decoding errors beyond half the minimum distance, we complement this paper by Monte–Carlo simulations on the *q-ary symmetric channel* (QSC), in which we compare the decoding performance of classical BMD decoding, the original Sudan algorithm [3] and Algorithm 2.

At first, we consider the code $\mathcal{RS}\left(2^8; 255, 63\right)$. A BMD decoder can correct up to 96 errors with this code. In contrast to this, Algorithm 2 is able to decode error patterns up to $t_{\max}^{[2]} = 107$. The same holds for the original Sudan algorithm [3]. Fig. 4 shows the QSC decoding performance for a BMD decoder, for Algorithm 2 and for the Sudan algorithm. If a transmitted symbol is received correctly with probability $1 - p = 0.7$, we observe that the word error rate obtained by Algorithm 2 is more than 100 times smaller than the word error rate obtainable by BMD decoding. The simulation results for the Sudan algorithm coincide very well with the simulation results obtained by Algorithm 2. Hence, our algorithm basically yields the same decoding performance as the classical Sudan algorithm. Moreover, we see that the bound calculated by (35) is only slightly worse than the actual word error performance of Algorithm 2, which means that Corollary 4 is well suited to get reliable estimates for the decoding performance.

Next, we consider the code $\mathcal{RS}\left(2^8; 255, 38\right)$, whose rate is approximately 0.15 and pursuant to (13) we have $l = 3$. For this case, we do not have an analytical bound on $P_w$. However, the simulation results depicted in Fig. 5 show that also for this case, the decoding performance of Algorithm 2 is basically identical to to the decoding performance of the Sudan algorithm. Hence, from a practical point of view, our decoding algorithm can compete well with the original Sudan algorithm [3]. Moreover, if the crossover probability of the QSC is approximately $p = 0.4$, that is, if the transmitted symbols are received correctly with probability $1 - p = 0.6$, Algorithm 2 achieves a word error rate which is more than 10000 times smaller than the word error rate obtainable by classical BMD decoding.

## VIII. CONCLUSIONS

In this paper, we have considered a syndrome based method for decoding low–rate Reed–Solomon codes beyond half the minimum distance. This is achieved by virtually extending the received Reed–Solomon words into words of an heterogeneous IRS code. In this way, we are able to use an efficient IRS decoding method described in [20], which is based on multi–sequence shift–register synthesis, which can be performed in the same order of complexity as the well–known Berlekamp–Massey algorithm.

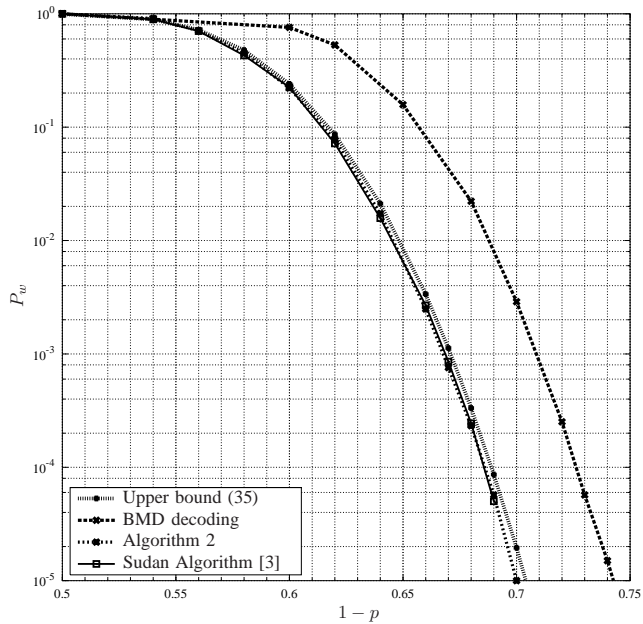

Fig. 4. Comparison of the QSC decoding performance of the Reed–Solomon code $\mathcal{RS}\left(2^8; 255, 63\right)$ for BMD decoding, Algorithm 2, and the Sudan Algorithm.

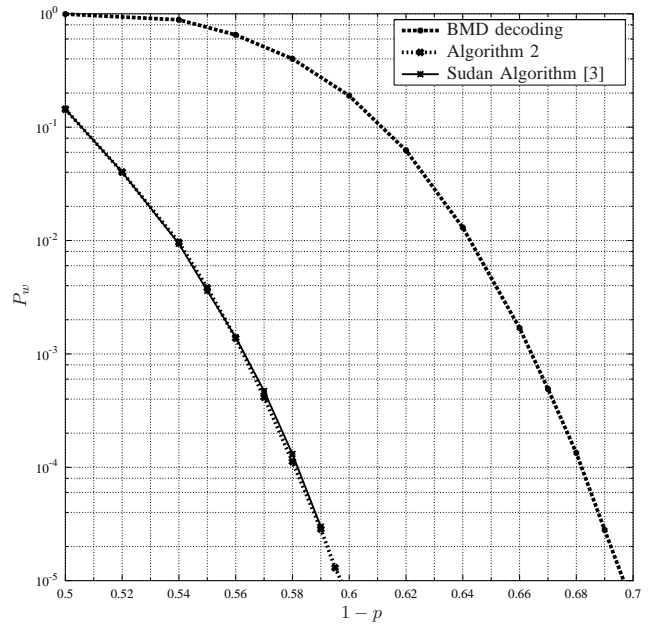

Fig. 5. Comparison of the QSC decoding performance of the Reed–Solomon code $\mathcal{RS}\left(2^8; 255, 38\right)$ for BMD decoding, Algorithm 2, and the Sudan Algorithm.

Most results derived in [20] for heterogeneous IRS codes are also valid for the "virtual" IRS codes obtained by syndrome extension. However, the result on the failure probability $P_w(t)$ stated in [20] cannot be applied to analyze syndrome extension decoding, since the erroneous columns cannot be considered to be uniformly distributed over all non-zero columns.

Nevertheless, even if the theoretical analysis of $P_w(t)$ is much more involved in the case of syndrome extension decoding, we have been able to derive and prove an upper bound on the failure probability for codes of rate $1/6 \lesssim R \lesssim 1/3$ defined over fields of characteristic two. This bound theoretically proves the effectiveness of the decoding procedure described by Algorithm 2. To check the tightness of this bound, we performed Monte–Carlo experiments for codes defined over the extension field $\mathbb{F}_{2^5}$, and showed that our bound coincides quite well with the experimental results.

Moreover, to compare our decoder with the original Sudan algorithm [3], and to get an impression of gains which can be achieved by our decoder, we performed Monte–Carlo simulations for the $q$-ary symmetric channel. We observed, that the performance obtained by Algorithm 2 is practically identical to the performance of the Sudan algorithm.

The decoding method presented in this paper can be implemented quite efficiently and can therefore be an interesting alternative to the Sudan algorithm. Unfortunately, like the original Sudan algorithm, its application is limited to codes of rate $R \lesssim 1/3$. For the Sudan algorithm there exists an extension by Guruswami and Sudan [11] which overcomes this problem. So, its remains a major challenge to extend the approach presented here to be able to decode Reed–Solomon codes of higher rates. Moreover, it would be desirable to find an universal bound on $P_w(t)$, which allows it to analytically evaluate the performance of Algorithm 2 for arbitrary Reed–Solomon codes.

## References

[1] R. E. Blahut, "Transform techniques for error control codes," *IBM J. Research and Development*, vol. 23, pp. 299–315, May 1979.
[2] S. Egonov, G. Markarian, and K. Pickavance, "A modified Blahut algorithm for decoding Reed–Solomon codes beyond half the minimum distance," *IEEE Trans. Commun.*, vol. 52, pp. 2052–2056, December 2004.
[3] M. Sudan, "Decoding of Reed-Solomon codes beyond the error correction bound," *J. Complexity*, vol. 13, pp. 180–193, March 1997.
[4] G. Schmidt, V. R. Sidorenko, and M. Bossert, "Decoding Reed–Solomon codes beyond half the minimum distance using shift-register synthesis," in *Proc. IEEE Intern. Symposium on Inf. Theory*, (Seattle, WA, USA), pp. 459–463, July 2006.
[5] W. W. Peterson, "Encoding and error–correction procedures for the Bose–Chaudhuri codes," *IEEE Trans. Inform. Theory*, vol. IT-6, pp. 459–470, September 1960.
[6] E. R. Berlekamp, *Algebraic Coding Theory*. New York: McGraw–Hill, 1968.
[7] J. L. Massey, "Shift-register synthesis and BCH decoding," *IEEE Trans. Inform. Theory*, vol. IT-15, pp. 122–127, January 1969.
[8] J. L. Massey and T. Schaub, "Linar complexity in coding theory," in *Coding Theory and Applications*, vol. 311 of *Lecture Notes in Computer Science*, pp. 19–32, Berlin: Springer Verlag, 1988.
[9] G. D. Forney, "On decoding BCH codes," *IEEE Trans. Inform. Theory*, vol. IT-11, pp. 549–557, October 1965.
[10] R. E. Blahut, *Theory and Practice of Error Control Codes*. Reading, Massachusetts: Addison-Wesley Publishing Company, 1983. ISBN 0-201-10102-5.
[11] V. Guruswami and M. Sudan, "Improved decoding of Reed-Solomon and algebraic-geometric codes," *IEEE Trans. Inform. Theory*, vol. IT-45, pp. 1755–1764, September 1999.

[12] R. Roth and G. Ruckenstein, "Efficient decoding of Reed–Solomon codes beyond half the minimum distance," *IEEE Trans. Inform. Theory*, vol. IT-46, pp. 246–257, January 2000.
[13] D. Coppersmith and M. Sudan, "Reconstructing curves in three (and higher) dimensional space from noisy data," in *Proc. ACM Symposium on Theory of Computing*, (San Diego, USA), pp. 136–142, June 2003.
[14] V. Y. Krachkovsky and Y. X. Lee, "Decoding for interleaved Reed-Solomon schemes," *Trans. Magn.*, vol. 33, pp. 2740–2743, September 1997.
[15] V. Y. Krachkovsky, Y. X. Lee, and H. K. Garg, "Decoding of parallel RS codes with applications to product and concatenated codes," in *Proc. IEEE Intern. Symposium on Inf. Theory*, (Boston, USA), p. 55, 1998.
[16] D. Bleichenbacher, A. Kiayias, and M. Yung, "Decoding of interleaved Reed Solomon codes over noisy data," in *Springer Lecture Notes in Computer Science*, vol. 2719, pp. 97–108, January 2003.
[17] A. Brown, L. Minder, and A. Shokrollahi, "Probabilistic decoding of interleaved RS-codes on the $q$-ary symmetric channel," in *Proc. IEEE Intern. Symposium on Inf. Theory*, (Chicago, IL, USA), p. 327, 2004.
[18] J. Justesen, C. Thommesen, and T. Høholdt, "Decoding of concatenated codes with interleaved outer codes," in *Proc. IEEE Intern. Symposium on Inf. Theory*, (Chicago, IL, USA), p. 329, 2004.
[19] G. Schmidt, V. R. Sidorenko, and M. Bossert, "Error and erasure correction of interleaved Reed–Solomon codes," in *Proc. Int. Workshop on Coding and Cryptography*, (Bergen, Norway), pp. 20–29, March 2005.
[20] G. Schmidt, V. R. Sidorenko, and M. Bossert, "Collaborative decoding of interleaved Reed–Solomon codes and concatenated code designs." Preprint, available online at ArXiv, `arXiv:cs.IT/0610074`, 2006.
[21] F. G. Gustavson, "Analysis of the Berlekamp-Massey linear feedback shift-register synthesis algorithm.," *IBM J. Research and Development*, vol. 20, pp. 204–212, May 1976.
[22] G.-L. Feng and K. K. Tzeng, "An iterative algorithm of shift-register synthesis for multiple sequences," *Scientia Sinica (Science in China), Series A*, vol. XXVIII, pp. 1222–1232, November 1985.
[23] G.-L. Feng and K. K. Tzeng, "A generalization of the Berlekamp-Massey algorithm for multisequence shift-register synthesis with applications to decoding cyclic codes," *IEEE Trans. Inform. Theory*, vol. IT-37, pp. 1274–1287, September 1991.
[24] G. Schmidt and V. R. Sidorenko, "Multi-sequence linear shift-register synthesis: The varying length case," in *Proc. IEEE Intern. Symposium on Inf. Theory*, (Seattle, WA, USA), pp. 1738–1742, July 2006.
[25] G. Schmidt and V. R. Sidorenko, "Linear shift-register synthesis for multiple sequences of varying length." Preprint, available online at ArXiv, `arXiv:cs.IT/0605044`, 2006.
[26] J. Feldman, M. J. Wainwright, and D. R. Karger, "Using linear programming to decode binary linear codes," *IEEE Trans. Inform. Theory*, vol. IT-51, pp. 954–972, March 2005.